\documentclass[12pt,preprint]{aastex}
\usepackage{color}

\shortauthors{Stanek et al.}

\shorttitle{Hypernova Bump in GRB~041006}

\begin{document}
\title{Deep Photometry of GRB~041006 Afterglow: Hypernova Bump at
Redshift $z=0.716$\altaffilmark{1}}

\author{K.~Z.~Stanek\altaffilmark{2},
P.~M.~Garnavich\altaffilmark{3},
P.~A.~Nutzman\altaffilmark{2},
J.~D.~Hartman\altaffilmark{2},
A.~Garg\altaffilmark{2},
K.~Adelberger\altaffilmark{4},
P.~Berlind\altaffilmark{5},
A.~Z.~Bonanos\altaffilmark{2},
M.~L.~Calkins\altaffilmark{5},
P.~Challis\altaffilmark{2},
B.~S.~Gaudi\altaffilmark{2},
M.~J.~Holman\altaffilmark{2},
R.~P.~Kirshner\altaffilmark{2},
B.~A.~McLeod\altaffilmark{2},
D.~Osip\altaffilmark{6},
T.~Pimenova\altaffilmark{3},
T. H. Reiprich\altaffilmark{7},
W.~Romanishin\altaffilmark{8},
T.~Spahr\altaffilmark{2},
S.~C.~Tegler\altaffilmark{9},
X.~Zhao\altaffilmark{3}
}

\altaffiltext{1}{Based on data from the MMTO 6.5m telescope, the 1.8m
Vatican Advanced Technology Telescope, the Magellan 6.5m Baade and
Clay telescopes, and the Keck II 10m telescope}

\altaffiltext{2}{\vspace{-0.15cm}Harvard-Smithsonian Center for Astrophysics, 60
Garden Street, Cambridge, MA 02138}
\altaffiltext{3}{\vspace{-0.15cm}Dept.~of Physics, University of Notre Dame, 225
Nieuwland Science Hall, Notre Dame, IN 46556}
\altaffiltext{4}{\vspace{-0.15cm}Carnegie Observatories, 813 Santa Barbara Street,
Pasadena, CA 91101}
\altaffiltext{5}{\vspace{-0.15cm}F.~L.~Whipple Observatory, 670 Mt.~Hopkins Road,
P.O.~Box 97, Amado, AZ 85645}
\altaffiltext{6}{\vspace{-0.15cm}Carnegie Institution of Washington,
Las Campanas Observatory, Casilla 601, La Serena, Chile}
\altaffiltext{7}{\vspace{-0.15cm}Institut f. Astrophysik u. Extr. Forschung,
Universit\"at Bonn, Auf dem H\"ugel 71, 53121 Bonn, Germany}
\altaffiltext{8}{\vspace{-0.15cm}Department of Physics and Astronomy, University of Oklahoma, 440 West Brooks, Norman, OK 73019}
\altaffiltext{9}{\vspace{-0.15cm}Physics \& Astronomy, Northern Arizona University,
Flagstaff, AZ 86011}

\email{\small 
kstanek@cfa.harvard.edu,
pgarnavi@miranda.phys.nd.edu,
pnutzman@cfa.harvard.edu,
jhartman@cfa.harvard.edu,
agarg@cfa.harvard.edu,
kurt@ociw.edu,
pberlind@cfa.harvard.edu,
abonanos@cfa.harvard.edu,
mcalkins@cfa.harvard.edu
pchallis@cfa.harvard.edu,
sgaudi@cfa.harvard.edu,
mholman@cfa.harvard.edu,
rkirshner@cfa.harvard.edu,
bmcleod@cfa.harvard.edu,
dosip@lco.cl,
tpimenov@nd.edu,
thomas@reiprich.net,
wjr@nhn.ou.edu,
tspahr@cfa.harvard.edu,
Stephen.Tegler@nau.edu,
xzhao@nd.edu
}

\begin{abstract}

We present deep optical photometry of the afterglow of gamma-ray burst
(GRB) 041006 and its associated hypernova obtained over $65\;$days
after detection (55 $R$-band epochs on 10 different nights).  Our
early data ($t<4\;$days) joined with published GCN data indicates a
steepening decay, approaching $F_\nu \propto t^{-0.6}$ at early times
($t\ll 1\;$day) and $F_\nu \propto t^{-1.3}$ at late times. The break
at $t_b=0.16\pm 0.04\;$days is the earliest reported jet break among
all GRB afterglows. During our first night, we obtained 39 exposures
spanning $2.15\;$hours from 0.62 to 0.71 days after the burst that
reveal a smooth afterglow, with an $rms$ deviation of $0.024\;$mag
from the local power-law fit, consistent with photometric errors.
After $t\sim 4\;$days, the decay slows considerably, and the light
curve remains approximately flat at $R\sim 24\;$mag for a month before
decaying by another magnitude to reach $R\sim 25\;$mag two months
after the burst. This ``bump'' is well-fitted by a k-corrected light
curve of SN1998bw, but only if stretched by $\times1.38$ in time.  In
comparison with the other GRB-related SNe bumps, GRB\,041006 stakes
out new parameter space for GRB/SNe, with a very bright and
significantly stretched late-time SN light curve.  Within a small
sample of fairly well observed GRB/SN bumps, we see a hint of a
possible correlation between their peak luminosity and their ``stretch
factor'', broadly similar to the well-studied Phillips relation for
the type Ia supernovae.

\end{abstract}


\section{Introduction}

At least some gamma-ray bursts (GRBs) are produced by events with the
spectra and light curves of core-collapse supernovae (SNe), as
demonstrated decisively by two GRBs which occurred in 2003,
GRB\,030329/SN\,2003dh \citep[e.g.][]{stanek03, Matheson03} and
GRB\,031203/SN\,2003lw \citep[e.g.][]{Malesani04}. At redshifts of
$z=0.1685$ and $z=0.1055$, respectively, these are classical GRBs with
the two lowest redshifts measured to date (although GRB\,031203 was
most likely sub-luminous compared to other classical GRBs).
GRB\,980425 at $z=0.008$ was also associated with ``hypernova'' 1998bw
\citep{Galama98}, but the isotropic energy of that burst was 10$^{-3}$
to 10$^{-4}$ times weaker than classical cosmological GRBs which might
place it in a unique class.

In addition, late-time deviations from the power-law decline typically
observed for optical afterglows have been seen in a number of cases
and these bumps in the light curves have been interpreted as evidence
for supernovae \citep[for a recent summary of their properties,
see][]{Zeh04}.  Possibly the best case of a supernova bump was
provided by GRB\,011121 which occurred at a redshift of $z = 0.36$ and
thus would have had a relatively bright supernova component.  A bump
in the light curve was observed both from the ground and with the
\emph{HST} \citep{Garnavich03, Bloom02}.  The color changes in the
light curve of GRB\,011121 were also consistent with a supernova,
designated SN~2001ke \citep{Garnavich03}.

Due to this mounting evidence, there have been claims that all
supernovae that produce GRBs are type~Ic hypernovae. The bias toward
type~Ic is partly due to the perceived difficulty in producing a jet
that escapes from a star with a massive envelope.  But this prejudice
may not be justified; after all, it was recently believed that due to
the large baryon content of supernovae it was not possible for any
supernova to be a GRB source.  It is therefore prudent to still assume
that the range of SN types that are responsible for GRBs, and their
properties, is an unsolved observational problem.  In several cases
there is evidence that GRBs could indeed be associated with other type
of supernovae, such as type IIn \citep{Garnavich03}, or normal type
Ib/c or fainter hypernovae \citep{price03, DellaValle03, Fynbo04}.
Obtaining magnitudes, colors and spectra of more GRB supernovae is
clearly a top priority in understanding the origin of long/soft
bursts.

GRB\,041006 was detected and localized by the French Gamma Ray
Telescope and the Wide Field X-Ray Monitor instruments aboard the
\emph{High Energy Transient Explorer II} at 12:18:08 (UT is used
throughout this paper) on 2004 October~6 \citep{Galassi04}.  It was
classified as a ``X-ray rich GRB'', and it was similar to GRB 030329
in its lightcurve shape and spectral characteristics.
\cite{DaCosta04} and \cite{Price04a} reported discovery of a new,
fading optical source within the $5.0\arcmin$ \emph{HETE} error
circle, located at $\alpha$~=~$00^{\rm h}54^{\rm m}50\fs2$,
$\delta$~=~$+01\arcdeg14\arcmin07\farcs$ (J2000.0), and identified it
as the GRB optical afterglow.  \cite{Fugazza04} and \cite{Price04b}
obtained an absorption line redshift of $z=0.716$ for the afterglow,
which is fairly low for a GRB. Motivated by the continued need to
study the GRB/SN connection as discussed above, we undertook
photometric monitoring of the optical afterglow of GRB\,041006.  In
this paper, we present photometry of the afterglow of GRB\,041006
obtained throughout the two months immediately following the burst,
which resulted in a clear detection of a light curve bump, most likely
due to the underlying hypernova. We discuss the properties of that
hypernova compared to the sample of other well studied GRB-related
SNe.

\begin{figure}
\plotone{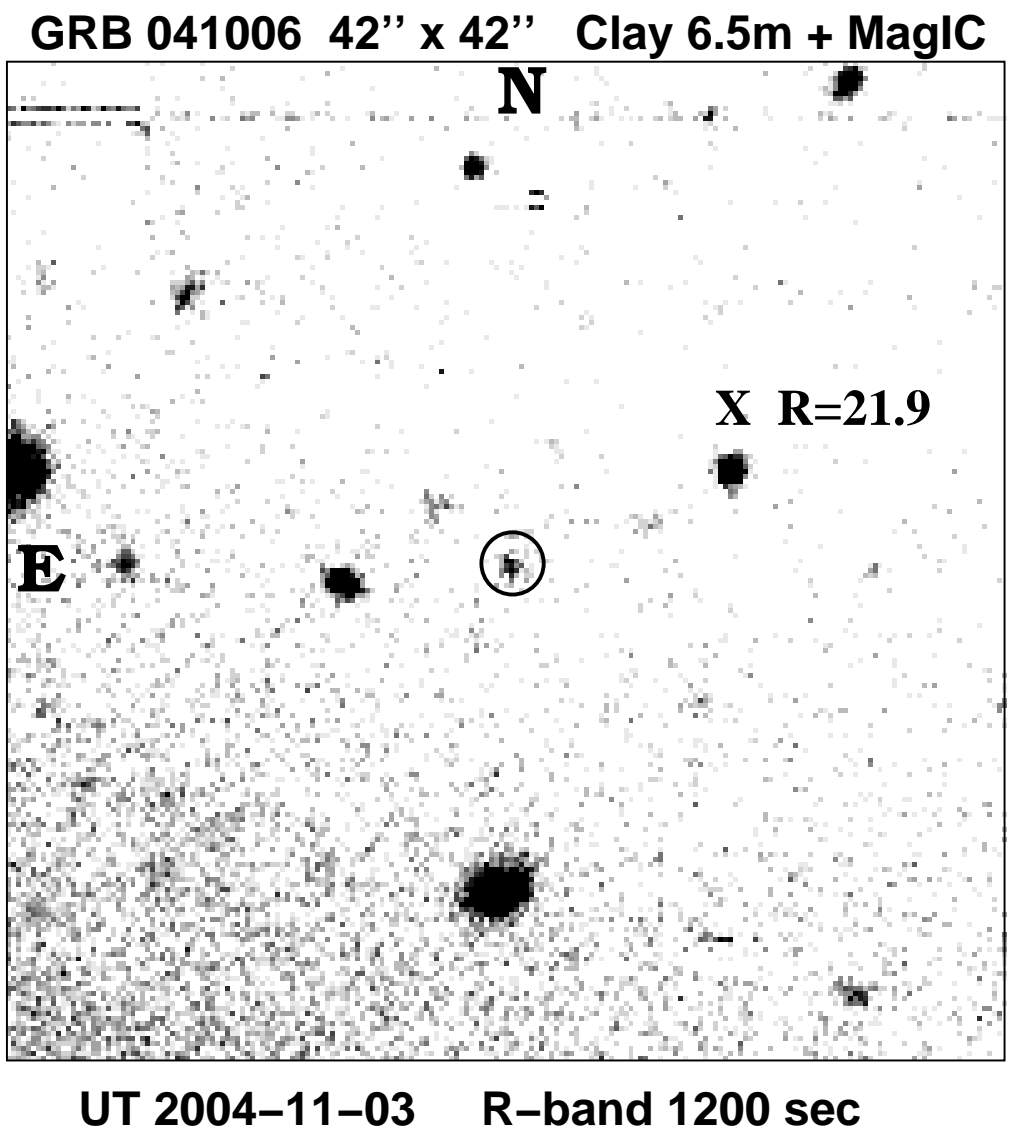}
\caption{$R$-band image of the afterglow ({\it circle}) of GRB\,041006
obtained with the Clay 6.5m telescope on UT Nov. 3, 2004 ($27.5\;$days
after the burst). The brightness of the afterglow is $R=24.08$. The
photometric comparison star is marked with an ``X''. {\em This figure
will not be included in the ApJL version of the published paper
due to space constraints.}}
\label{fig_findingchart}
\end{figure}

\section{The Photometric Data}

The photometric data we obtained are listed in
Table~\ref{phottable}\footnote{The analysis presented here supersedes
our GCN Circulars by \citet{Garnavich04b} and \citet{Garg04}.}.  The
majority of our data were obtained with the MegaCam CCD mosaic
\citep{mcleod00} mounted on the MMT 6.5m telescope. Additional data
were obtained using the 1.8m Vatican Advanced Technology Telescope,
the Magellan 6.5m Baade and Clay telescopes, and the Keck II 10m
telescope\footnote{All photometry presented in this paper are
available through {\tt anonymous ftp} on {\tt cfa-ftp.harvard.edu}, in
the directory {\tt pub/kstanek/GRB041006}. Images are available by
request.}.

All the data were reduced using DAOPHOT II
\citep{stetson87,stetson92}.  For the MMT data, we have used 10-50
stars to establish the zero-point offset between frames. We found that
a nearby star, marked ``X'' in Figure~\ref{fig_findingchart}, is
constant over many days to within $0.02\;$mag, with $R=21.90$ using
the calibration of \citet{Garnavich04b} and \citet{Henden04}. For
other data, we use star ``X'' as the photometric calibrator. For the
MegaCam data, where a Sloan $r'$ filter was used, we found that it
translates well into standard $R_C$ filter magnitudes without a need
for a color term.

\section{Temporal Behavior in the Photometry}

The transition between the phase dominated by the afterglow and the
phase dominated by the supernova is marked by an abrupt change in the
behavior of the light curve.  In \S \ref{earlyphot} we consider
photometry before this observed change, which occurs about $4.5\;$days
after the burst.  In \S \ref{laterphot} we consider data taken during
the following two months.

\begin{figure}
\plotone{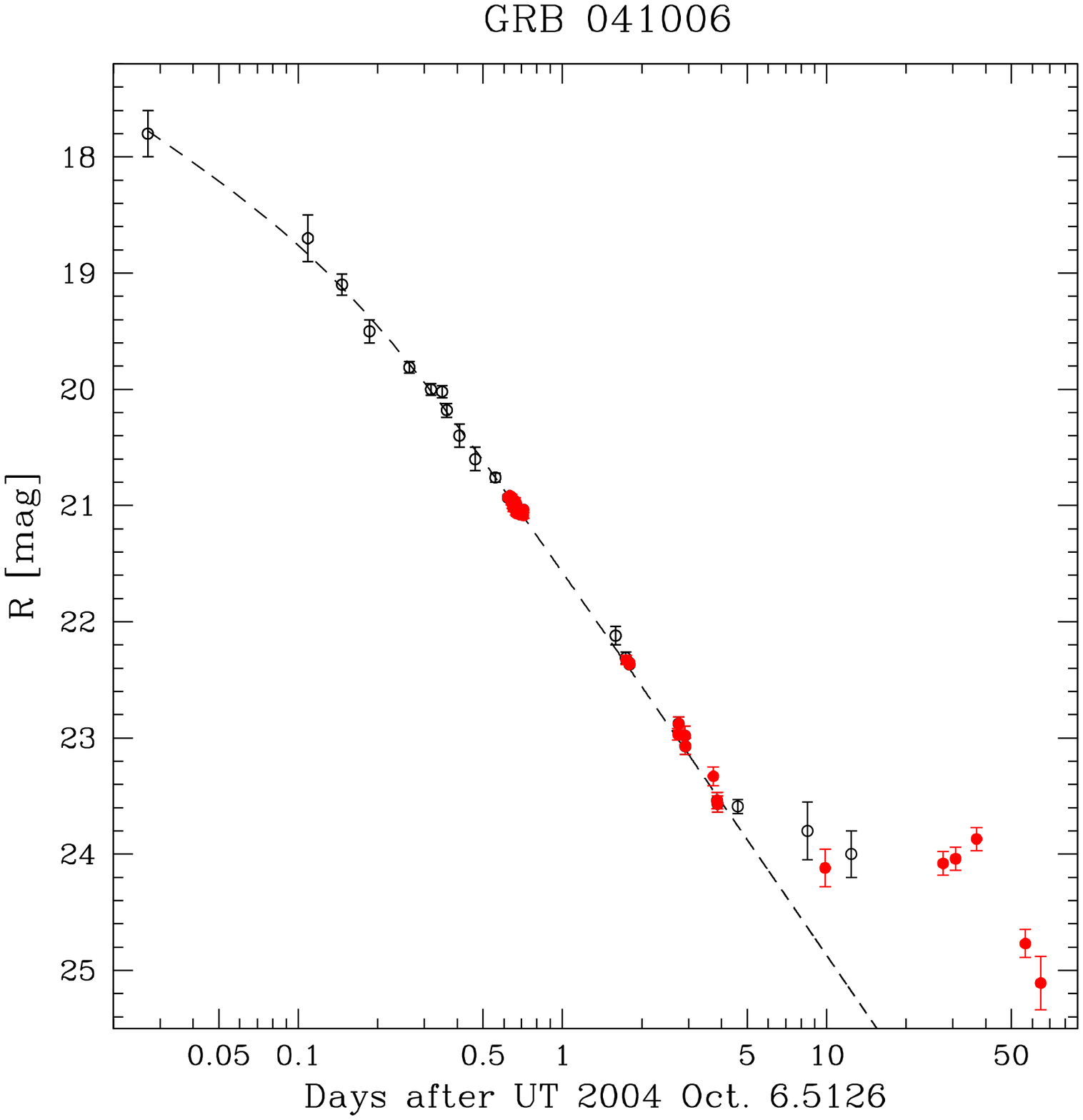}
\caption{$R$-band light curve for GRB\,041006.  Early data (days 1-4)
are fit to a broken power-law {\it (dashed curve)}.  The fit yields a
very early break time, $t_b=0.14$~days, which might be the earliest
jet break reported to date.  Filled points are our data, while open
circles are data from GCN (\citealt{Kinugasa04}; \citealt{Kahharov04};
\citealt{Fugazza04}; \citealt{DAvanzo04}; \citealt{Monfardini04};
\citealt{Misra04}; \citealt{Covino04}; \citealt{Balman04};
\citealt{Bikmaev04}).  }
\label{fig_lightcurve}
\end{figure}

\subsection{Early Photometry: Days 1-4}
\label{earlyphot}

Figure~\ref{fig_lightcurve} presents the $R$-band light curve for
GRB\,041006. We have added some data published via the GCN (see the
caption of Figure~\ref{fig_lightcurve} for the references) to fill in
the gaps in our data. All of the GCN data were brought to the common
zero-point, and if no photometric error was reported, we adopted a
conservative value of $0.2\;$mag. We have fit the GRB\,041006 data
with the broken power-law model of \citet{Beuermann99} (see also
\citealt{Stanek01}):
\begin{equation}
F_{R}(t) =
\frac{2F_{R,0}}{\left[\left(\frac{t}{t_b}\right)^{\alpha_1 s}
+\left(\frac{t}{t_b}\right)^{\alpha_2 s}\right]^{1/s}},
\label{eq_brokenpower}
\end{equation}
where $t_b$ is the time of the break, $F_{R,0}$ is the $R$-band flux
at $t_b$ and $s$ is a parameter which determines the sharpness of the
break, where a larger $s$ gives a sharper break.  This formula
smoothly connects the early time $t^{-\alpha_1}$ decay ($t\ll t_b$)
with the later time $t^{-\alpha_2}$ decay ($t\gg t_b$). Equation
\ref{eq_brokenpower} has been used to describe the afterglow decay of
e.g. GRBs 990510 (with $s=1$) and 010222 \citep{Stanek99, Stanek01}.
The fit results in the following values for the parameters:
$\alpha_1=0.59\pm 0.13,\alpha_2=1.32\pm 0.02, t_b=0.16\pm 0.04$~days
(we have fixed the value of $s=2.5$). Given the non-uniform data used
(i.e. GCN plus our data), the quoted values for the errors of these
parameters should be treated with caution.

The resulting fit is shown as the dashed curve in Figure
\ref{fig_lightcurve}.  The curve is a good fit to the early afterglow,
with $\chi^2/DOF= 1.1$. The fit yields a very early break time,
$t_b=0.16\pm 0.04$~days, which is the earliest reported afterglow
break (see the compilation of break times, $t_{jet}$ in Table 1 of
\citealt{Friedman05}; see also Ghirlanda et al.~2004).  Given the
implications for the jet opening angle and the energetics of the
burst, it would be valuable to further constrain the break time with
robust calibration of other early data.

Short-timescale variability has been seen in the light curves of
several afterglows, including GRB\,011211 \citep[e.g.][]{Holland02}
and GRB\,021004 \citep[e.g.][]{Bersier03}.  Motivated to look for
such short-timescale variability in GRB\,041006, the first night on
the MMT we obtained 39 high S/N exposures of the afterglow, spanning
2.15 hours from 0.62 to 0.71 days after the burst.  No short-timescale
variability is present in our data, and the rms deviation of
$0.024\;$mag from the local power-law fit is consistent with random
errors of the photometry. The afterglow of GRB\,041006 joins other
well observed afterglows, such as GRB\,990510 \citep{Stanek99} and
GRB\,020813 \citep{Laursen03} in the category of very smooth
afterglows.

\subsection{Later Photometry: Days 5-65}
\label{laterphot}

\begin{figure}
\plotone{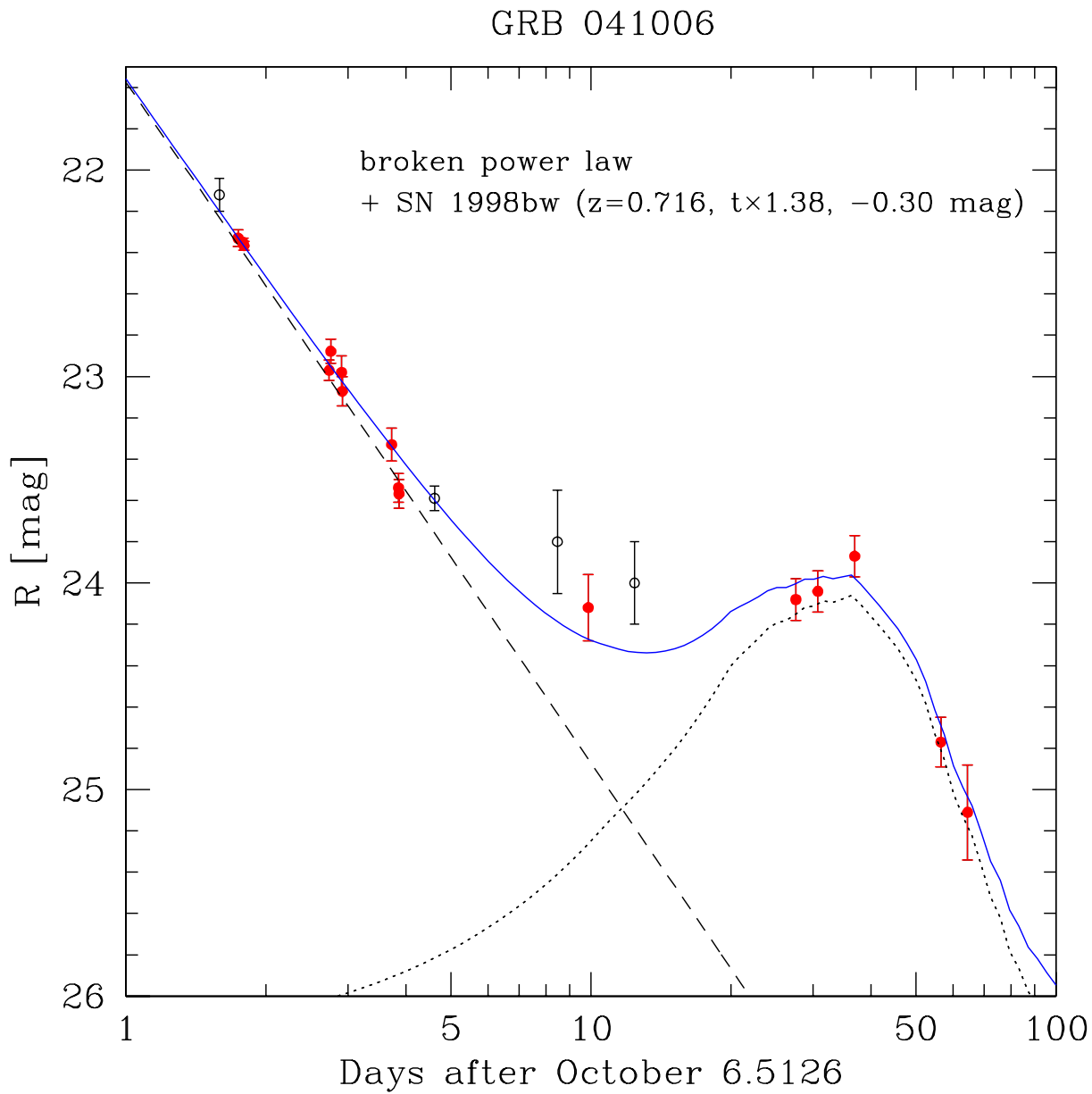}
\caption{Late $R$-band light curve for GRB\,041006.  Open circles are
data from GCN, while the filled points show our data.  The early-time
broken power-law decay of the optical afterglow is shown as the dashed
line. A k-corrected SN\,1998bw light curve extended by a factor of 1.38
is shown as the dotted curve.  The combined power law and stretched SN
1998bw light curve is given by the solid curve. }
\label{fig_later}
\end{figure}

Over the subsequent two months we obtained $R$-band observations of
the OT on a number of epochs.  We combine our data with three
additional $R$-band photometric measurements published via GCN
(\citealt{Covino04}; \citealt{Balman04}; \citealt{Bikmaev04}). These
observations are presented in Figure~\ref{fig_later}.

As is apparent in Figure~\ref{fig_later}, the decay abruptly slows
$\sim 5$ days after the burst.  For slightly more than a month, the
brightness then remains roughly flat at $R \sim 24$ mags.  Over the
next 30 days, the OT decays by another magnitude, falling to $R\sim25$
mag in our final observations.

This clear detection of a light curve ``bump'' strongly suggests the
late-time dominance of the OT by an underlying supernova component.
The supernova component peaks at roughly $\sim24\;$mag about
$35\;$days after the burst, which is several magnitudes brighter than
the extrapolated optical afterglow of the GRB at the time of this
peak.

To estimate the properties of the hypernova bump in GRB\,041006, we
correct for the effects of Galactic extinction using the reddening map
of \citet{sfd98}.  At the Galactic coordinates of GRB\,041006,
$l=124\arcdeg\!\!.74, b=-61\arcdeg\!\!.66$, the foreground reddening
is $E(B-V)=0.023$, yielding an expected extinction of $A_R=0.06$.  We
find that the OT light curve can be well modeled by a stretched
SN\,1998bw-like bump, k-corrected to redshift $z=0.716$, combined with
the continued power-law decay of the GRB optical afterglow.  It is
clear from the data that the amount of stretching of the SN\,1998bw
light curve needed is significant, and it is basically determined by
the rapid decline by about a magnitude from $36.9$ to $56.5\;$days
after the burst.  The composition is shown as the solid line in
Figure~\ref{fig_later}. Indeed, we find a good fit using as a template
light curve of SN 1998bw, uniformly stretched in time by a factor of
$1.38\pm 0.06$ (in addition to the $1+z$ cosmological time dilation),
and with significantly brighter peak absolute magnitude ($0.3\;$mag
brighter). We did not fit for the brightness of the host, as its
influence on the derived $R$-band magnitudes would vary for our data
set obtained with different telescopes, instruments and varying seeing
conditions.  To account for that, we adopt an asymmetric error bar for
the peak brightness of the supernova of $-0.1,+0.2\;$mag.

Thus the hypernova component in the late-time OT of GRB\,041006 is
long-lasting, in contrast to e.g. GRB\,01121/SN\,2001ke
\citep{Garnavich03} and GRB\,030329/SN\,2003dh \citep{Lipkin04,
deng05}, in which the bumps decay faster than SN\,1998bw (see next
Section for more discussion).  As in the case of
GRB\,030329/SN\,2003dh \citep{Matheson03}, we find no need to
introduce a time delay (either positive or negative) between the
afterglow and the supernova component to fit the light curve data. In
fact, adding a time delay parameter degrades the $\chi^2/DOF$ value
slightly.  The best fit time delay is $-0.9\pm 2.7\;$days, i.e. the
GRB and the SN were most likely simultaneous.

\section{Summary and Discussion}

We have obtained deep photometry of the optical transient associated
with GRB\,041006 for the first two months following the burst.  We
find that the early-time $R$-band light curve is well fit by a broken
power law, with a pre-break index of $\alpha_{1}=0.59$ and a steeper
post-break index of $\alpha_{2}=1.32$.  Our fit yields a break time of
$t_{b}=0.16$ days after the burst, the earliest jet break observed so
far.  Early observations also show a very smooth afterglow, with
short-timescale deviations from the power-law decay on the same order
($0.025\;$mag) as random errors in photometry.

Continuing observations of the optical transient over the two months
following the burst show an abrupt change from the $t^{-1.3}$ decay of
the afterglow, consistent with the domination of the photometry by a
SN component.  The SN light curve peaks at a similar brightness to SN
1998bw but takes more time following the burst to evolve, by a factor
of $\sim 1.38$.

\begin{figure}
\plotone{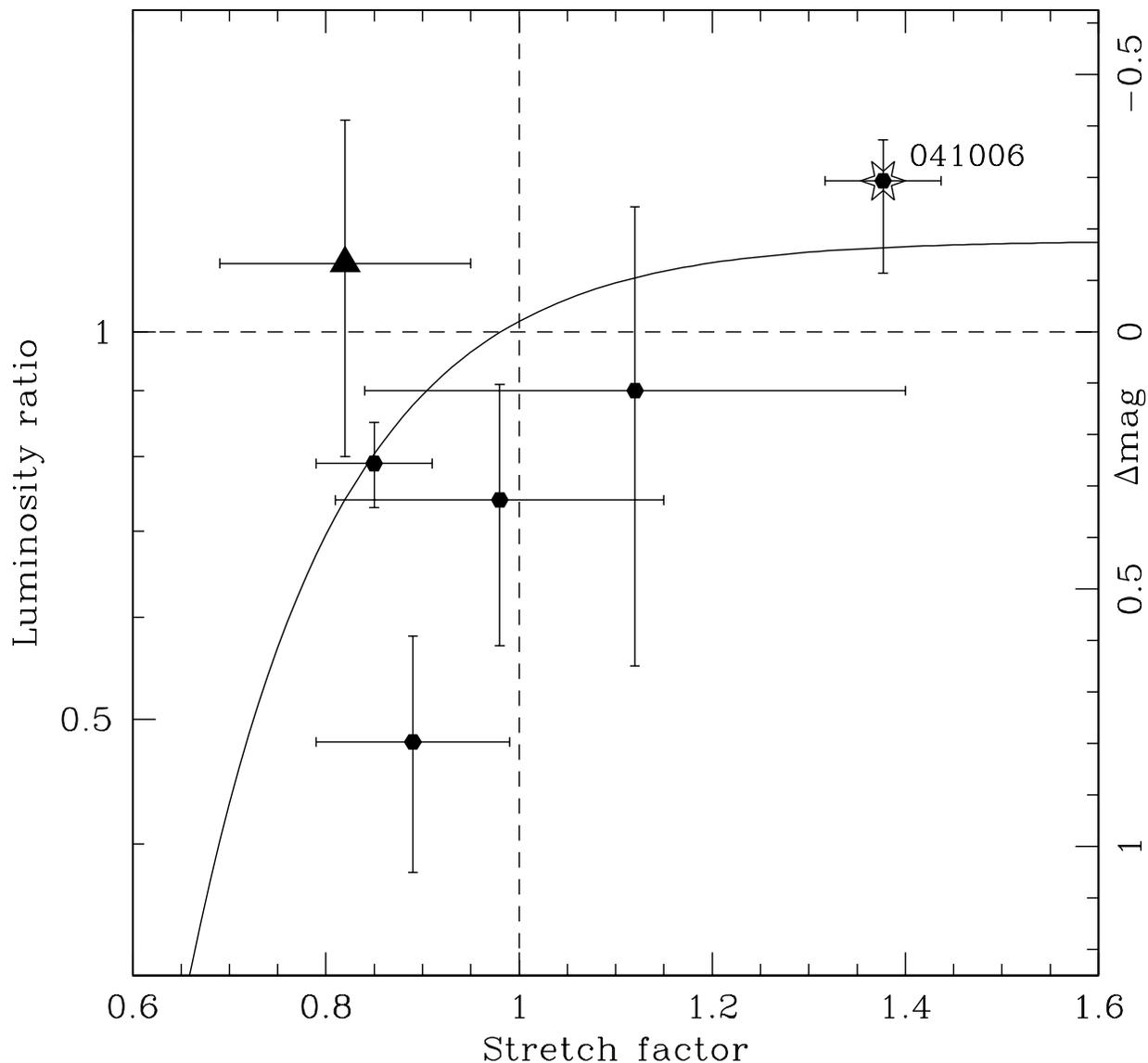}
\caption{Time stretch factor relative to SN 1998bw versus luminosity
relative to SN\,1998bw for six GRB/SNe (see also Figure~3 from ZKH04).
Dashed lines at stretch factor $= 1$ and luminosity ratio $= 1$
represent SN\,1998bw values.  The four filled circles represent data
for four GRB/SNe given in ZKH04.  GRB\,030329/SN\,2003dh ({\it
triangle}) and the SN component for GRB\,041006 ({\it star}) are also
shown (see text for more discussion). Plotted with the continuous line
is the analogous relation for type Ia supernovae, adopted from
\citet{Garnavich04a}. }
\label{fig_zeh}
\end{figure}

To put the light curve features of GRB\,041006 into the context of
previous GRB/SN observations, we have adopted Figure~3 from
\citealt{Zeh04} (hereafter ZKH04), which presents the deduced time
stretch factors relative to SN 1998bw versus their deduced luminosity
ratios of GRB/SNe relative to SN 1998bw.  They consider nine GRBs,
observed before the end of 2002, for which there exists some evidence
of a SN bump.  They have suggested a subdivision of these GRBs into a
group of five with weak evidence for a bump, and a group of four with
statistically significant evidence for a bump (GRBs 990712, 991208,
011121, and 020405; see \S 3 of ZKH04).  In our Figure \ref{fig_zeh},
we present only this latter group, with the additions of
GRB\,030329/SN\,2003dh, which occurred after their cutoff date, and
GRB\,041006. For GRB\,030329/SN\,2003dh, we adopt the values for the
parameters from a recent paper by \citealt{Zeh05} (hereafter
ZKH05).  It should be stressed that given the bumpy and long-lasting
afterglow of GRB\,030329, the exact parameters for the underlying
SN\,2003dh are still subject to debate (\citealt{Matheson03, Lipkin04,
deng05}). We decided not to include another recent GRB/SN case,
GRB\,031203/SN\,2003lw, as it was heavily obscured by foreground
Galactic dust and resided in a rather luminous host galaxy.  For that
supernova, ZKH05 give values for the stretch of $1.14\pm 0.16$, and
luminosity ratio of $1.65\pm 0.41$.  We note that, among this sample
of six, GRB\,041006 possesses the highest redshift, with
$z=0.717$. Also very interestingly, \emph{all} GRBs with optical
afterglows and measured redshifts less than that of GRB\,041006 show
evidence of late-time SN components (ZKH04).  As apparent in
Figure~\ref{fig_zeh}, GRB\,041006 stakes out new parameter space for
GRB/SNe, with a very bright and significantly stretched late-time SN
light curve.

The distribution of light curve stretch factors and peak brightnesses
for hypernovae shown in Figure~\ref{fig_zeh} is reminiscent of the
relation found by Phillips (1993) for type~Ia SNe (technique
introduced by Perlmutter et al.~1997 for their sample of type Ia). We
overplot the type~Ia relation extended to fast declining events by
Garnavich et al.~(2004a) and extrapolated to large stretch
factors. While the uncertainties for the hypernovae make a conclusive
statement difficult, we note that both the hypernovae and type~Ia
events are likely powered by 56Ni decay and have similar peak
luminosities implying comparable 56Ni masses. But with the present
data available for them, hypernovae are still a long way from being
useful distance indicators.  More events like GRB\,041006 will be
possible to study soon, as the {\em Swift} satellite \citep{gehrels04}
is already providing accurate localizations for a good number of GRBs.

\acknowledgments

We would like to thank the staffs of the MMT, VATT, Las Campanas and
Keck Observatories.  We thank the anonymous referee for useful
comments. We are grateful to Chris Stubbs for his help in obtaining
some of the data. We thank Bohdan Paczy\'nski for reading an earlier
version of the manuscript and for suggesting comparison with type Ia
SNe in Fig.4. We thank the \emph{HETE2} team, Scott Barthelmy, and the
GRB Coordinates Network (GCN) for the quick turnaround in providing
precise GRB positions to the astronomical community, and we thank all
the observers who provided their data and analysis through the GCN.
PMG acknowledges the support of NASA/LTSA grant NAG5-9364. JDH is
funded by a National Science Foundation Graduate Student Research
Fellowship. BSG is supported by a Menzel Fellowship from the Harvard
College Observatory. THR acknowledges the F.H. Levinson Fund of the
Peninsula Community Foundation.

\begin{deluxetable}{ r c c c c }
\tabletypesize{\small}
\tablewidth{0pt}
\tablecaption{JOURNAL OF PHOTOMETRIC OBSERVATIONS\label{phottable}}
\tablehead{\colhead{$\Delta$T\tablenotemark{a}} &
\colhead{$R_C$} &
\colhead{$\sigma_R$} &
\colhead{$t_{exp}$ (s)} &
\colhead{Observatory\tablenotemark{b}}
}
\startdata
 0.6225 & 20.931 & 0.023 & 120 & MMT \\ 
 0.6311 & 20.918 & 0.017 & 180 & MMT \\ 
 0.6397 & 20.973 & 0.021 & 120 & MMT \\ 
 0.6418 & 20.925 & 0.021 & 120 & MMT \\ 
 0.6437 & 20.954 & 0.020 & 120 & MMT \\ 
 0.6455 & 20.952 & 0.028 & 120 & MMT \\ 
 0.6473 & 20.997 & 0.027 & 120 & MMT \\ 
 0.6492 & 20.995 & 0.033 & 120 & MMT \\ 
 0.6510 & 21.023 & 0.030 & 120 & MMT \\ 
 0.6528 & 20.989 & 0.023 & 120 & MMT \\ 
 0.6547 & 20.975 & 0.035 & 120 & MMT \\ 
 0.6565 & 21.005 & 0.022 & 120 & MMT \\ 
 0.6583 & 20.990 & 0.019 & 120 & MMT \\ 
 0.6601 & 21.010 & 0.027 & 120 & MMT \\ 
 0.6620 & 20.962 & 0.030 & 120 & MMT \\ 
 0.6639 & 20.996 & 0.038 & 120 & MMT \\ 
 0.6657 & 21.051 & 0.032 & 120 & MMT \\ 
 0.6675 & 20.991 & 0.021 & 120 & MMT \\ 
 0.6693 & 21.023 & 0.021 & 120 & MMT \\ 
 0.6712 & 21.070 & 0.023 & 120 & MMT \\ 
 0.6730 & 21.020 & 0.026 & 120 & MMT \\ 
 0.6749 & 21.037 & 0.026 & 120 & MMT \\ 
 0.6767 & 21.057 & 0.025 & 120 & MMT \\ 
 0.6785 & 21.028 & 0.019 & 120 & MMT \\ 
 0.6804 & 21.028 & 0.028 & 120 & MMT \\ 
 0.6822 & 21.073 & 0.029 & 120 & MMT \\ 
 0.6840 & 21.056 & 0.030 & 120 & MMT \\ 
 0.6859 & 21.051 & 0.030 & 120 & MMT \\ 
 0.6877 & 21.036 & 0.028 & 120 & MMT \\ 
 0.6895 & 21.052 & 0.025 & 120 & MMT \\ 
 0.6913 & 21.034 & 0.026 & 120 & MMT \\ 
 0.6931 & 21.077 & 0.025 & 120 & MMT \\ 
 0.7014 & 21.079 & 0.024 & 120 & MMT \\ 
 0.7032 & 21.050 & 0.019 & 120 & MMT \\ 
 0.7051 & 21.044 & 0.023 & 120 & MMT \\ 
 0.7069 & 21.047 & 0.021 & 120 & MMT \\ 
 0.7087 & 21.052 & 0.020 & 120 & MMT \\ 
 0.7105 & 21.036 & 0.022 & 120 & MMT \\ 
 0.7124 & 21.086 & 0.025 & 120 & MMT \\ 
 1.7407 & 22.330 & 0.04  & 1500 & VATT \\ 
 1.7851 & 22.351 & 0.020 & 600  & MMT \\ 
 1.7931 & 22.367 & 0.022 & 600  & MMT \\ 
 2.7341 & 22.970 & 0.05  & 4800  & VATT \\ 
 2.7557 & 22.878 & 0.06  & 600  & MMT \\ 
 2.9086 & 22.980 & 0.08  & 600  & MMT \\ 
 2.9176 & 23.072 & 0.07  & 600  & MMT \\ 
 3.7266 & 23.330 & 0.08  & 4200  & VATT \\ 
 3.8531 & 23.539 & 0.07  & 600  & MMT  \\ 
 3.8613 & 23.568 & 0.07  & 600  & MMT \\ 
 9.8622 & 24.119 & 0.16  & 600  & MMT  \\ 
27.5763 & 24.080 & 0.10  & 1200 & Clay \\ 
30.7511 & 24.040 & 0.10  & 1200 & Clay \\ 
36.9040 & 23.870 & 0.11  & 720  & Keck II\\ 
56.5242 & 24.770 & 0.12  & 1800 & Baade \\ 
64.5516 & 25.110 & 0.23  & 1200 & Baade 
\enddata
\tablenotetext{a}{Days after 2004 October 6.5126 UT.}
\tablenotetext{b}{MMT: MMT telescope/MegaCam;
VATT: Vatican Advanced Technology Telescope;
Clay: Magellan Clay tel./MagIC;
Keck II: Keck II tel./ESI;
Baade: Magellan Baade tel./IMACS
}
\end{deluxetable}


\begin{thebibliography}{}

\bibitem[Balman et al.(2004)]{Balman04} Balman, S., et al. 2004, GCN
Circ.~2821

\bibitem[Bersier et al.(2003)]{Bersier03} Bersier, D., et al. 2003,
\apj, 584, L43

\bibitem[Beuermann et al.(1999)]{Beuermann99} Beuermann, K., et al.
1999, A\&A, 352, 26

\bibitem[Bikmaev et al.(2004)]{Bikmaev04} Bikmaev, I., et al. 2004,
GCN Circ.~2826

\bibitem[Bloom et~al.(2002)]{Bloom02} Bloom, J.~S., et~al., 2002,
ApJ, 572, L45

\bibitem[Covino et al.(2004)]{Covino04} Covino, S., et al. 2004,
GCN Circ.~2803

\bibitem[D'Avanzo et al.(2004)]{DAvanzo04} D'Avanzo, P. et al. 2004,
GCN Circ.~2788

\bibitem[Da Costa, Noel, \& Price(2004)]{DaCosta04} da Costa, G.~S.,
Noel, N., \& Price, P.~A. 2004, GCN Circ.~2765

\bibitem[Della Valle et~al.(2003)]{DellaValle03} Della Valle, M., et 
al.\ 2003, A\&A, 406, L33 

\bibitem[Deng et al.(2005)]{deng05} Deng, J., Tominaga, N., Mazzali,
P. A., Maeda, K., \& Nomoto, K. 2005, ApJ, in press (astro-ph/0501670)

\bibitem[Friedman \& Bloom(2005)]{Friedman05} Friedman, A.~S. \&
Bloom, J.~S. 2005, \apj, submitted (astro-ph/04084113)

\bibitem[Fugazza et al.(2004)]{Fugazza04} Fugazza, D., et al. 2004,
GCN Circ.~2782

\bibitem[Fynbo et~al.(2004)]{Fynbo04} Fynbo, J. P. U., et al. 2004, 
ApJ, 609, 962

\bibitem[Galassi et al.(2004)]{Galassi04} Galassi, M.,~et al. 
2004, GCN Circ.~2770

\bibitem[Galama et al.(1998)]{Galama98} Galama, T.~J.,~et al. 
1998a, \nat, 395, 670 

\bibitem[Garg et al.(2004)]{Garg04} Garg, A., Stubbs, C., Challis, P., 
Stanek, K.~Z.; Garnavich, P. 2004, GCN Circ.~2829

\bibitem[Garnavich et al.(2003)]{Garnavich03} Garnavich, P.~M., et
al. 2003, \apj, 582, 924

\bibitem[Garnavich et al.(2004a)]{Garnavich04a} Garnavich, P.~M., et
al. 2004a, \apj, 613, 1120

\bibitem[Garnavich et al.(2004b)]{Garnavich04b} Garnavich, P., Zhao, X.,
\& Pimenova, T. 2004b, GCN Circ.~2792

\bibitem[Gehrels et~al.(2004)]{gehrels04} Gehrels, N. et~al.
2004, ApJ, 611, 1005
	
\bibitem[Ghirlanda et al.(2004)]{ghirlanda04} Ghirlanda, G., Ghisellini, G.,\&  
Lazzati, D. 2004, ApJ, 616, 331 

\bibitem[Henden(2004)]{Henden04} Henden, A. A. 2004, GCN Circ.~2801

\bibitem[Holland et~al.(2002)]{Holland02} Holland, S.~T., et al., 
2002, AJ, 124,639

\bibitem[Kahharov et al.(2004)]{Kahharov04} Kahharov, B., Asfandiyarov, I.,
Ibrahimov, M., Sharapov, D., Pozanenko, A.,  Rumyantsev, V., \& Beskin, G.
2004, GCN Circ.~2775

\bibitem[Kinugasa \& Torii(2004)]{Kinugasa04} Kinugasa, K.  \& Torii, K. 2004,
GCN Circ.~2832

\bibitem[Laursen \& Stanek(2003)]{Laursen03} Laursen, L. T. \& Stanek, K. Z,
2003, \apj, 597, L107

\bibitem[Lipkin et~al.(2004)]{Lipkin04} Lipkin, Y. M., et~al.
2004, \apj, 606, 381

\bibitem[Malesani et~al.(2004)]{Malesani04} Malesani, D., et~al. 
2004, \apj, 609, L5 

\bibitem[Matheson et al.(2003)]{Matheson03} Matheson, T.,~et al.
2003, \apj, 599, 394

\bibitem[McLeod et al.(2000)]{mcleod00} 
McLeod, B. A., Conroy, M., Gauron, T. M., Geary, J. C., \& Ordway, M. P.
2000, fdso.conf,~11

\bibitem[Misra \& Pandey(2004)]{Misra04} Misra, K. \& Pandey, S.~B. 2004, 
GCN Circ.~2794

\bibitem[Monfardini et al.(2004)]{Monfardini04} Monfardini, A., et al. 2004,
GCN Circ.~2790

\bibitem[Phillips (1993)]{phillips93} Phillips, M. M. 1993, \apj, 413, L105 

\bibitem[Perlmutter et al.(1997)]{Perlmutter97} Perlmutter et al. 1997,
ApJ, 483, 565


\bibitem[Price et al.(2003)]{price03} Price, P.~A., et al.~2003, 
ApJ, 589, 838

\bibitem[Price, Da Costa, \& Noel(2004)]{Price04a} Price, P.~A., da Costa,
G.~S, \& Noel, N. 2004, GCN Circ.~2771

\bibitem[Price et al.(2004)]{Price04b} Price, P.~A.,~et al. 2004, 
GCN Circ.~2791

\bibitem[Schlegel, Finkbeiner, \& Davis(1998)]{sfd98} Schlegel, D.~J.,
Finkbeiner, D.~P., \& Davis, M.\ 1998, \apj, 500, 525

\bibitem[Stanek et al.(1999)]{Stanek99} Stanek, K.~Z., Garnavich, P.~M., 
Kaluzny, J., Pych, W., Thompson, I. 1999, \apj, 522, L39

\bibitem[Stanek et al.(2001)]{Stanek01} Stanek, K.~Z., et al. 2001,
\apj, 563, 592

\bibitem[Stanek et al.(2003)]{stanek03} Stanek, K.~Z., et al. 2003,
\apj, 591, L17

\bibitem[Stetson(1987)]{stetson87} Stetson, P.~B. 1987, \pasp, 99 191

\bibitem[Stetson(1992)]{stetson92} Stetson, P.~B. 1992, in ASP
Conf.~Ser.~25, Astrophysical Data Analysis Software and Systems I,
ed. ~D.~M.~Worrall, C.~Bimesderfer, \& J.~Barnes (San Francisco: ASP),
297

\bibitem[Zeh, Klose, \& Hartmann(2004)]{Zeh04} Zeh, A., Klose, S. \& Hartmann,
D.H.\ 2004, \apj, 609, 952

\bibitem[Zeh, Klose, \& Hartmann(2005)]{Zeh05} Zeh, A., Klose, S. \&
Hartmann, D.H.\ 2005, to appear in Proc. "22nd Texas Symposium on
Relativistic Astrophysics" (astro-ph/0503311)

\end{thebibliography}
\end{document}